\documentclass[cits]{PoS}
\usepackage[mathscr]{euscript}
\usepackage{graphicx}
\usepackage{amsmath}

\def\chpt{\raise0.4ex\hbox{$\chi$}PT}
\def\schpt{S\raise0.4ex\hbox{$\chi$}PT}
\def\rschpt{rS\raise0.4ex\hbox{$\chi$}PT}
\def\figref#1{Fig.~\ref{fig:#1}}
\def\Figref#1{Figure~\ref{fig:#1}}

\def\tabref#1{Table~\ref{tab:#1}}

\def\gtwid{{\,\raise.3ex\hbox{$>$\kern-.75em\lower1ex\hbox{$\sim$}}\,}}
\def\ltwid{{\,\raise.3ex\hbox{$<$\kern-.75em\lower1ex\hbox{$\sim$}}\,}}

\def\ie{{\it i.e.},\ }

\def\et{{\it et al.}}

\def\vs{{\it vs.}\ }

\def\rcite#1{Ref.~\cite{#1}}

\def\eqn#1{\label{eq:#1}}

\def\eq#1{Eq.~(\ref{eq:#1})}

\def\eqs#1#2{Eqs.~(\ref{eq:#1}) and (\ref{eq:#2})}

\title{Finite-volume effects and the electromagnetic contributions to kaon and pion masses}

\ShortTitle{Electromagnetic contributions to kaon and pion masses}

\author{S.~Basak$^a$, A.~Bazavov$^b$, \speaker{C.~Bernard}$^c$, C.~DeTar$^d$, E.~Freeland$^e$, 
 J.~Foley$^d$, Steven~Gottlieb$^f$, U.M.~Heller$^g$, J.~Komijani$^c$, J.~Laiho$^{h}$, L.~Levkova$^{d}$,
 J.~Osborn$^i$, R.L.~Sugar$^j$, A.~Torok$^f$\thanks{current address: Intel Corporation, Hillsboro, OR  97124, USA}\;, D.~Toussaint$^k$, R.S.~Van~de~Water$^{l}$, R.~Zhou$^l$ \\

$^a$ NISER, Bhubaneswar, Orissa 751005, India      \\
$^b$ Department of Physics and Astronomy,
University of Iowa, Iowa City, IA 52240, USA \\
$^c$ Department of Physics, Washington University, St. Louis, MO 63130, USA\\
$^d$ Department of Physics and Astronomy, University of Utah, Salt Lake City, UT 84112, USA\\
$^e$ Liberal Arts Department, School of the Art Institute of Chicago, Chicago, IL, USA \\ 
$^f$ Department of Physics, Indiana University, Bloomington, IN 47405, USA\\
$^g$ American Physical Society, One Research Road, Ridge, NY 11961, USA\\
$^h$ Department of Physics, Syracuse University, Syracuse, NY  13244, USA\\
$^i$ ALCF, Argonne National Laboratory, Argonne, IL 60439, USA\\
$^j$ Physics Department, University of California, Santa Barbara, CA 93106, USA\\
$^k$ Physics Department, University of Arizona Tucson, AZ 85721, USA\\
$^l$ Theoretical Physics Department, Fermi National Accelerator Laboratory, Batavia 60510, USA

\vspace{2mm}
{\bf MILC Collaboration}
\vspace{3mm}

E-mail: \email{cb@wustl.edu}
}
\abstract{
We report on the MILC Collaboration calculation of electromagnetic
effects on light pseudoscalar mesons.  The simulations employ asqtad
staggered dynamical quarks in QCD plus quenched photons, with lattice
spacings varying from 0.12 to 0.06 fm.  Finite volume corrections for
the MILC realization of lattice electrodynamics have been calculated
in chiral perturbation theory  and applied to the lattice data.
These corrections differ from those calculated by Hayakawa and Uno
because our treatment of zero modes differs from theirs.
Updated results for the corrections to ``Dashen's theorem'' are presented.
}

\FullConference{The 32nd International Symposium on Lattice Field Theory \\
                23--28 June, 2014\\
                Columbia University, New York, NY}

\begin{document}

{\it Introduction. --}
The up and down quark masses may be calculated on the lattice from the masses of the $K^+$ and $K^0$. To do so, one
must remove the effects of electromagnetism (EM) from the kaon system. 
In lattice determinations of the phenomenologically  important quantity
$m_u/m_d$, both by us \cite{qrat,Bazavov:2014wgs} and by other groups \cite{Aoki:2013ldr},
the uncertainty in the EM contributions to kaon masses is the largest  source of error. 
We have been working on reducing these uncertainties  by calculating the EM effects directly on the lattice; 
progress has been reported previously in Refs.~\cite{basak,EM12}. 

Electromagnetism affects 
the calculation of $m_u /m_d$ from kaon masses primarily through the mass splitting
$(M^2_{K^\pm}-M^2_{K^0})^\gamma$, where 
$\gamma$ denotes the EM contribution, \ie the difference between this quantity in the
real world and in a world where all quark charges are set to zero (keeping renormalized quark masses
unchanged).
Dashen's theorem \cite{dashen} states that this kaon mass splitting is equal
to the 
pion splitting
$(M^2_{\pi^\pm}-M^2_{\pi^0})^\gamma$
at lowest order (LO) in chiral perturbation theory (\chpt).   However,  the corrections
to the LO relation are not small.
One may parameterize them by \cite{Aoki:2013ldr}
\vspace{-2mm}
\begin{equation}
\eqn{eps-def}
 (M^2_{K^\pm}-M^2_{K^0})^\gamma =(1+\epsilon) (M^2_{\pi^\pm}-M^2_{\pi^0})^{\textrm{expt}}\ ,
 \vspace{-2mm}
\end{equation}
where the experimental pion splitting is used, rather than the pion EM splitting.  The two are equal up to isospin
violating effects, which are NNLO in the pion system and therefore small.

Here, we describe our calculation of $\epsilon$  on the lattice, using
unquenched QCD but quenched photons.  
The EM-quenching effects on $\epsilon$ may be calculated 
and corrected to NLO in SU(3) \chpt, with controlled errors \cite{BD}.   
We calculate $\epsilon$ by computing $(M^2_{K^\pm}-M^2_{K^0})^\gamma$, and either using  $(M^2_{\pi^\pm}-M^2_{\pi^0})^{\textrm{expt}}$ or
$(M^2_{\pi^\pm}-M^2_{\pi^0})^\gamma$ in the right-hand side of \eq{eps-def}. 
Determining the mass of the true $\pi^0$ is costly,  however, since it has disconnected
EM diagrams even in the isospin limit.  Instead, we drop the disconnected diagrams, which are expected to be 
small, and simply find the RMS average mass of $u\bar u$ and $d\bar d$ mesons. We call
the pion obtained in this manner the ``$\pi^0$.'' Both the true $(M^2_{\pi^0})^\gamma$ and our  $(M^2_{\rm{``}\pi^0\rm{"}})^\gamma$ are 
small
since EM contributions to neutral mesons vanish in the chiral limit. 

A key unresolved issue in our previous
work was the size of the finite volume (FV) effects.  Comparisons of results on two lattice volumes showed
unexpectedly small differences \cite{EM12}, making it difficult to determine the true size of the
effect.  Here, we examine the FV effects in more detail, and show that they can be predicted with good
accuracy in one-loop (staggered) chiral perturbation theory. We can therefore correct for FV effects,
with a small residual systematic error.  After doing so we are able to quote a new result for the
 parameter $\epsilon$, and give corresponding results for $m_u/m_d$.

{\it Lattice setup. --}
We calculate meson masses on the (2+1)-flavor MILC asqtad ensembles, with quenched photon fields, and with lattice spacings ranging from  
$\approx\!0.12\;{\rm fm}$ to $\approx\!0.06\;{\rm fm}$.  \tabref{ensembles} shows the ensembles employed.  
The valence quarks have charges $\pm2/3 e$, $\pm1/3 e$ or 0,
where $e= e_{\rm phys}$ or (on most ensembles) $2e_{\rm phys}$ or $3e_{\rm phys}$, with $e_{\rm phys}\approx 0.303$ the physical electron charge.
\begin{table}
\begin{center}
\begin{small}
\begin{tabular}{|c|l|l|c|c|c|c|}
\hline\hline
$\approx a$[fm]& Volume
& $\beta$
& $m'_l/m'_s$& \# configs.  &$L$ (fm) & $m_\pi L$ \\
\hline\hline
0.12 & \hspace{2mm}$12^3\times64^*$ & 6.76& 0.01/0.05&  1000  & 1.4 & 2.7    \\
     & \hspace{2mm}$16^3\times64^*$ & 6.76& 0.01/0.05&  1303  & 1.8 & 3.6   \\
     & \hspace{2mm}$20^3\times64$ & 6.76& 0.01/0.05&  2254 & 2.3 &  4.5  \\
      & \hspace{2mm}$28^3\times64$ & 6.76& 0.01/0.05& \phantom{2}274 & 3.2& 6.3   \\
      & \hspace{2mm}$40^3\times64^*$ & 6.76& 0.01/0.05& \phantom{2}115 & 4.6& 9.0   \\
      & \hspace{2mm}$48^3\times64^*$ & 6.76& 0.01/0.05& 132+52 & 5.4& 10.8   \\
      & \hspace{2mm}$20^3\times64$ & 6.76& 0.007/0.05& 1261 & 2.3& 3.8   \\
      & \hspace{2mm}$24^3\times64$ & 6.76& 0.005/0.05& 2099 & 2.7& 3.8   \\
\hline
0.09  & \hspace{2mm}$28^3\times96$ &7.09& 0.0062/0.031& 1930 & 2.3&  4.1  \\
      & \hspace{2mm}$40^3\times96$ & 7.08& 0.0031/0.031& 1015 & 3.3& 4.2   \\
\hline
0.06  & \hspace{2mm}$48^3\times144$ & 7.47& 0.0036/0.018& \phantom{2}670 & 2.8& 4.5   \\
\hline\hline
\end{tabular}
\end{small}
\caption{Parameters of the (2+1)-flavor asqtad ensembles used in this study. Volumes marked with $*$ are currently used in the FV studies but not
in the full analysis. The quark masses $m'_l$ and $m'_s$ are the light and strange dynamical masses used in the runs.  The  number of configurations listed  as `132+52' for the $a\!\approx\!0.12\:$fm, $48^3\times 64$ ensemble gives values for
two independent streams, the first in single precision, and the second in double.  At the moment, we treat them as separate data, and do not 
average the results. The $40^3\times64$ and $48^3\times64$ $a\!\approx\!0.12\:$fm ensembles are new since the conference and are still being analyzed.\label{tab:ensembles} }
\vspace{-8mm}
\end{center}
\end{table}

{\it QED in Finite Volume. --}
With the non-compact realization of QED on the lattice, which we use, it is necessary to drop some zero-modes in a finite volume in order to have 
a convergent path integral. In particular, the action in Coulomb gauge for the zero component of the vector potential, $A_0$, is
$\frac{1}{2}\int\left(\partial_i A_0\right)^2$.  Since the $A_0$ mode with spatial momentum $\vec k=0$ has vanishing action, it must be dropped.
Similarly, the action for the spatial components $A_i$ is $\frac{1}{2}\int\left[\left(\partial_0 A_i\right)^2+\left(\partial_j A_i\right)^2\right]$.  Here
only the mode with 4-momentum $k_\mu=0$ must be dropped, and that is what we do.  
Hayakawa and Uno, in their calculation of EM FV effects in \chpt \cite{Hayakawa:2008an}, drop all  $A_i$ modes with $\vec k=0$.  This means that
the FV effects in the MILC calculations are different from those computed in \rcite{Hayakawa:2008an}.

To make explicit the difference between our set-up and that of Ref.~\cite{Hayakawa:2008an}, we give the spatial components
of the photon propagator in each case:
\vspace{-1mm}
\begin{eqnarray}
\langle A_i (k) A_j(-k)\rangle &= 
 &\begin{cases}\frac{1}{k^2}\left(\delta_{ij}-\frac{k_ik_j}{\vec k^2}\right), & \vec k \not=0;\\
0\ , & \vec k=0.
\end{cases} \hspace{22mm}\textrm{[Hayakawa-Uno]} \eqn{H-U-prop}\\
\langle A_i (k) A_j(-k)\rangle &=
 &\begin{cases}
\frac{1}{k^2}
 \left(\delta_{ij}-\frac{k_ik_j}{\vec k^2}\right), & \vec k \not=0; \\
\frac {1}{k^2}\delta_{ij}\ , & \vec k=0,\ k_0\not=0;\hspace{10mm}\textrm{[MILC]}\\
0\ , & \vec k=0,\ k_0=0.\vspace{-2.5mm}
\end{cases}\eqn{MILC-prop}
\end{eqnarray}
The propagator of the time component $A_0$ is the same in both cases, and is simply $1/\vec k^2$ for $\vec k\not=0$, and 0 for $\vec k=0$.
 The violation of Gauss's Law induced by the absence of the $\vec k=0$ $A_0$ mode makes it possible to have net charges on a FV torus with periodic boundary 
 conditions \cite{Hayakawa:2008an}.  But Gauss's Law has no implications for the spatial modes $A_i$, so does not distinguish 
 between \eqs{H-U-prop}{MILC-prop}.
 
 We note that the BMW Collaboration  \cite{Borsanyi:2014jba}  has recently calculated the FV effects for the QED in \eqs{H-U-prop}{MILC-prop}, 
 which they call QED$_{L}$ and QED$_{TL}$, respectively.  As part of their study of the proton-neutron mass difference in dynamical QCD+QED,
 they have independently found the key FV results described below, and have also worked out explicit asymptotic forms in large volumes.  

{\it Finite Volume Effects in Chiral Perturbation Theory. --}
The EM effects on pseudoscalar meson masses have been calculated to NLO in staggered chiral perturbation theory (\schpt) \cite{schpt}.
The calculation gives the splitting in squared mass, $\Delta M^2_{xy}\equiv M^2_{xy}- M^2_{x'y'}$,
between a Goldstone (taste $\xi_5$) meson made from valence quarks $x$ and $y$ with charges
$q_x$, $q_y$ and masses $m_x$, $m_y$, and the corresponding meson made  from quarks $x'$ and $y'$  with the same masses
but the quark charges set to zero.  The explicit formula for  $\Delta M^2_{xy}$  in terms of NLO chiral logarithms and 
low-energy  constants (LECs) can be found in Refs.~\cite{EM12,schpt}.

The logarithms in the chiral expressions come from three Feynman diagrams.  Two of them,  the sunset graph and the 
photon tadpole, have internal low-energy photons, while one, the meson tadpole, has only an internal meson line. The FV effects coming 
from the meson tadpole are completely standard (see 
\rcite{Bernard:2001yj} for the explicit formulas we use), and are in fact tiny for all our ensembles.  On the other hand, one should expect 
large FV effects from the photon graphs, even on the largest of our ensembles, since the photon is massless.

Although the photon tadpole diagram vanishes in dimensional regularization in infinite volume (IV), since there are no dimensional parameters in the 
integral, it is important in FV.  In fact, we need to add the tadpole to the sunset diagram in order for the FV difference (FV result minus 
IV result) to be finite in Coulomb gauge.  Once the photon diagrams are added, we can simply perform a brute-force computation of 
the FV difference.  We do this using the importance-sampling integration program VEGAS \cite{VEGAS}.  The VEGAS integrand is taken to be the
difference between the IV integrand, and the distance-weighted average of its evaluations at the 16 corners of the FV hypercube containing the point.  
We have checked that our result for the sum of the sunset and the photon tadpole diagrams agrees with that of \rcite{Hayakawa:2008an} when \eq{H-U-prop} is used.
  
The difference between the Hayakawa and Uno result and the MILC result is simply due to the modes $\vec k=0$, $k_0\not=0$ in \eq
{MILC-prop}.  This difference 
can easily be worked out analytically, and is $qT/4L^3$, where $q$ is the meson charge, and $T$ and $L$ are the time and spatial length of the lattice, respectively.
Note that this gives a rather subtle large-volume behavior in the MILC case:  The result is acceptable if the limit $L\to\infty$ is taken before
$T\to\infty$, or if the limits are taken together at fixed aspect ratio $T/L$, but not if the limit $T\to\infty$ is taken first.  In other words, the 
MILC set-up is not well defined in finite spatial volume at zero temperature.  This fact has also been pointed out by
BMW \cite{Borsanyi:2014jba}.  They make the further point  that the QED$_{TL}$ (MILC) set-up violates reflection positivity because the 
constraint required to set the single $k_\mu=0$ mode of $A_i$ to zero involves the square of the integral over all space-time of $A_i$.  
Although many actions used in lattice QCD violate reflection positivity,  one might worry that in this case the violation leads to problems 
with defining or isolating the lowest states in correlation functions.  In practice, this does not seem to be a problem for us. We find no
discernible differences between the qualities of plateaus in correlation functions in our QCD+quenched QED simulations versus those 
for QCD alone. 

\begin{figure}
\vspace{-5mm}
\begin{center}\includegraphics[width=0.55\textwidth]{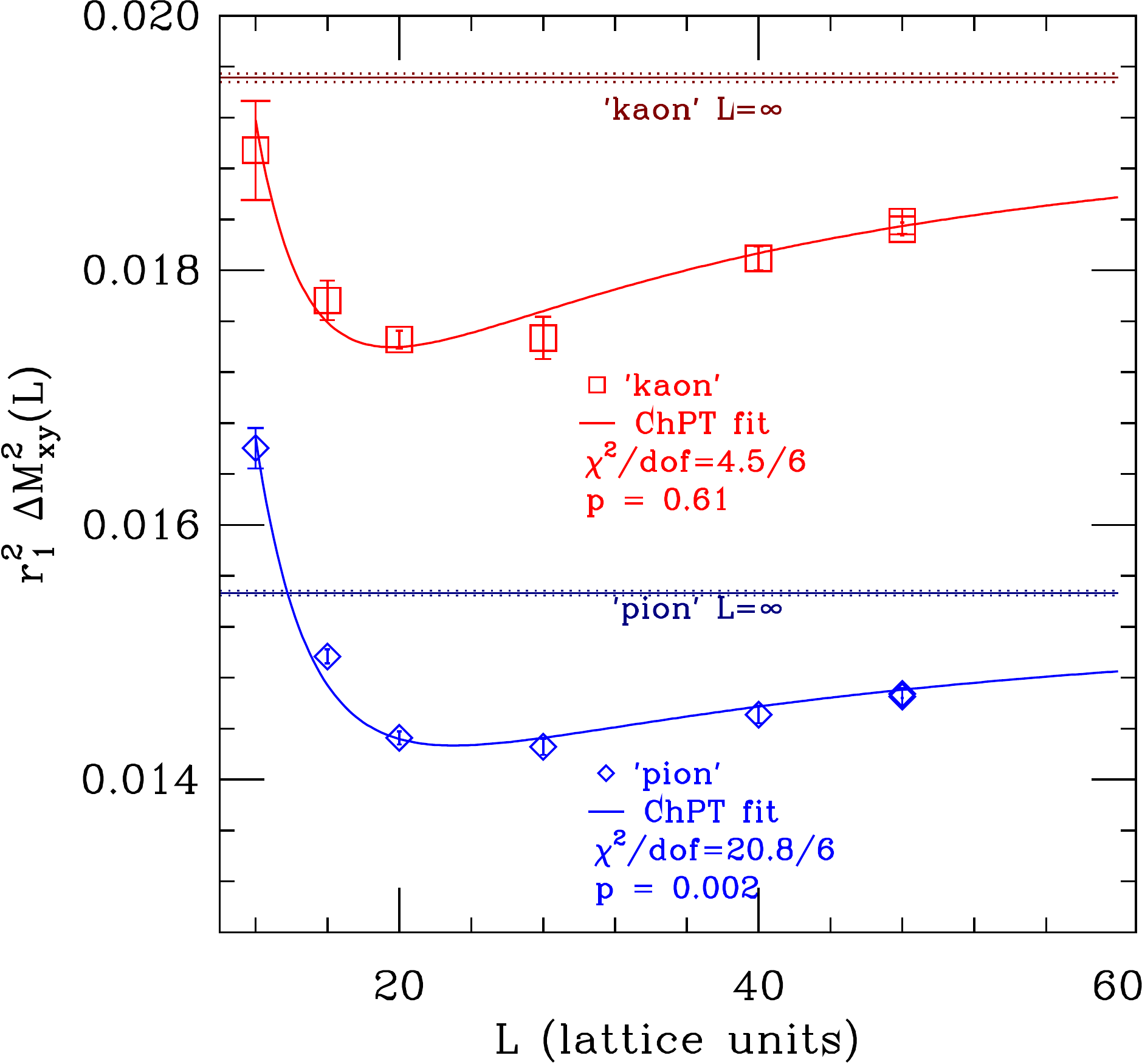}\end{center}
\vspace{-5mm}
\caption{\label{fig:FV} Finite volume effects at $a\approx0.12$ fm  and $am'_l=0.01, am'_s=0.05$ as a function of spatial lattice length $L$ for
two different meson masses: a unitary `pion' (blue) with degenerate valence masses $m_x=m_y=m'_l$, and a `kaon' (red) with valence masses
$m_x=m'_l$ and $am_y=0.04$, close to the physical strange quark mass.  The fit lines are to the FV form from \schpt, and have one free parameter each, the infinite volume value (shown by horizontal solid lines with dotted lines for errors).
}
\vspace{-0.15in}
\end{figure}

In \figref{FV} we show fits to the FV form derived in  \schpt\ for two different meson masses on the $a\!\approx\!0.12$ fm ensembles with 
$am'_l=0.01$, $am'_s=0.05$.  The shape of the fit curves are completely determined in NLO \schpt; the only free parameter 
in each fit is the overall 
height of the curve.  The theory at NLO gives a reasonably good description of the data, and we use it to correct the data for FV effects.
To estimate the systematic error associated with this correction (a `residual' FV error), we examine the deviations of the fit lines from the data in \figref{FV}. By far the largest deviation 
occurs for the  `pion' (blue) curve at $L=16$.  Although the fit there misses the point by $\approx\!4\sigma$, the difference between the data and 
the IV value is 70\% of the FV correction implied by the fit.  Below, we therefore take the residual FV  error to be 
30\% of the correction.  Since the deviation at all other points in \figref{FV} is much less than 30\%, we believe this error estimate to be  
conservative.  Note that a correction from higher order terms in \chpt\ of $\sim\!30\%$ is somewhat large, but not unreasonable.

One can now understand why it was difficult to observe FV effects directly in the data set available in \rcite{EM12}.  At that 
time, we had only the $L=20$ and $L=28$ ensembles to compare.  From \figref{FV}, one sees that the minima of the curves are  
in this region of $L$ or close to it, and therefore the difference expected between these volumes is small compared to the statistical errors in the data.

{\it Chiral Fits. --}
Once FV effects have been removed, we fit the data to the IV NLO \schpt\ form, plus NNLO analytic terms, as described in \cite{EM12}.  The higher order analytic terms, which include discretization terms, are necessary because our statistical errors in $\Delta M^2_{xy}$ are
$\sim\!0.3\%$ for charged mesons and $\sim\!1.0\%$ for neutral mesons. 
When we include data with charges greater than physical, analytical terms of order $\alpha^2_{EM}$ are also necessary to
obtain acceptable fits.  

\begin{figure}
\vspace{-5mm}
\begin{center}\includegraphics[width=0.65\textwidth]{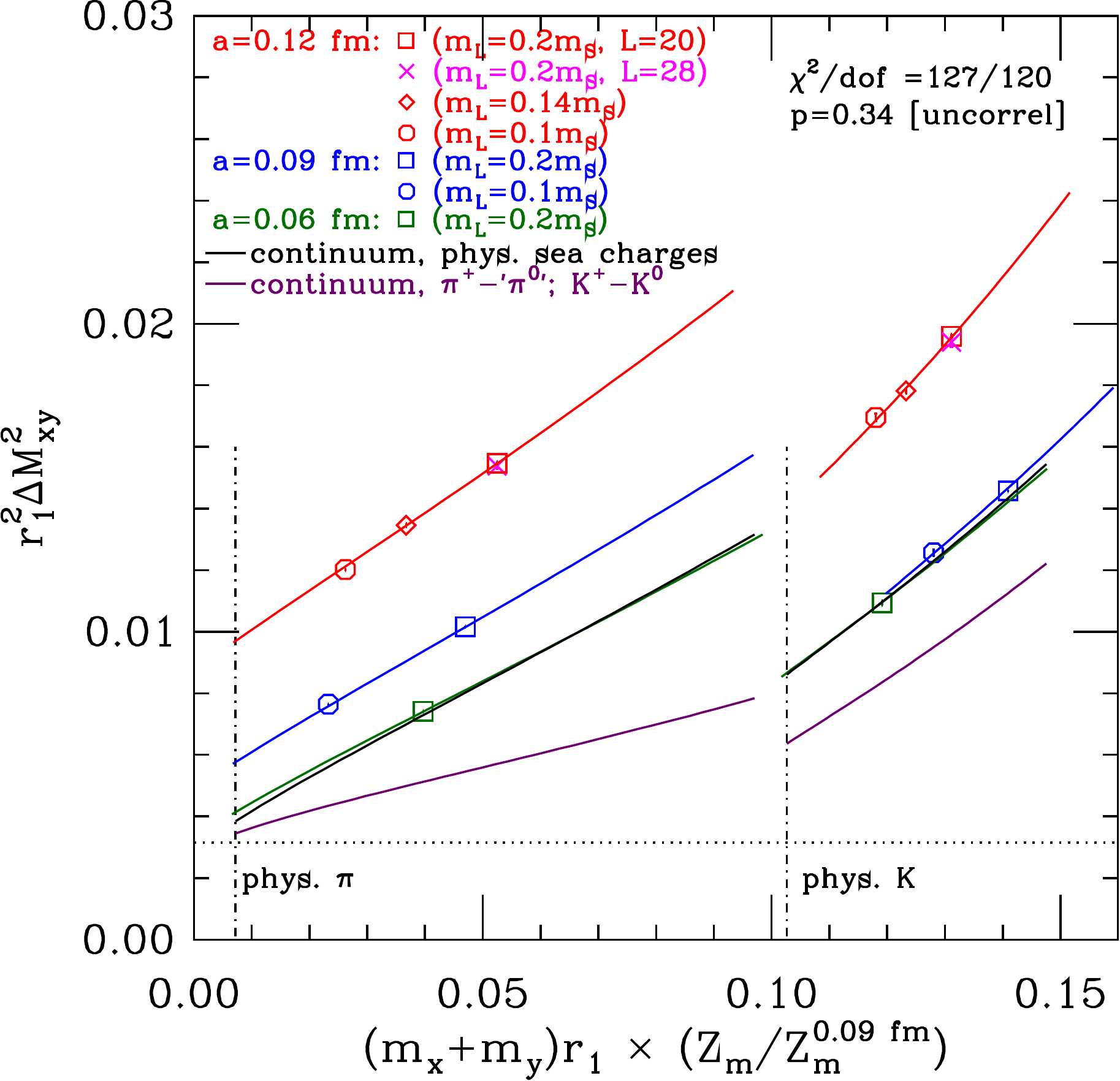}\end{center}
\vspace{-6mm}
\caption{Central fit to  the EM splitting $\Delta M_{xy}^2$ \vs the
sum of the valence-quark masses. Only a small subset of the partially quenched data set included in the fit is shown:  the points for $a\!\approx\!0.09$  fm and 
$\!\approx\!0.06$ fm, as well as the `pion' (left hand) points for $a\!\approx0.12$ fm, have unitary values of the valence masses, while the
`kaon' (right hand) points for $a\!\approx0.12$ fm have $m_x=m'_l$ but $m_y=0.8m'_s$, which is closer to the physical strange mass than
$m'_s$ itself.   All points shown are for mesons with charge $\pm e_{\rm phys}$; neutral mesons and mesons with charges 
$\pm 2e_{\rm phys}$ are also included in the fit.
The data have been corrected for FV effects using NLO \schpt.
The red, blue and green curves correspond to the three  lattice spacings.
The black and purple curves are extrapolations, see text.
The horizontal dotted line is the experimental value of the  $\pi^+$--$\pi^{0}$
splitting.
 \label{fig:EMfit}
\vspace{-4mm}
}
\end{figure}

\Figref{EMfit} shows our data after correction for FV effects, along with the chiral fit and its extrapolations.  The FV corrections 
are roughly 7--10\% for `pions' (shown on 
the left) and 10--18\% for `kaons' (shown on the right).  The corrections are larger at higher mass because of the overall factor of $M^2$ in the 
one-loop logarithms that give the FV effects, but not in the LO term, which is mass independent by Dashen's theorem.  
Partially quenched charged- and neutral-meson data, with  $e=\pm e_{\rm phys}$  and $e=\pm 2e_{\rm phys}$, are fit simultaneously, 
but only data for unitary or approximately unitary mesons  with charge $\pm e_{\rm phys}$ is shown in \figref{EMfit}.  This fit has 149 data points 
and 29 parameters, with $\chi^2=127$, for an (uncorrelated) $p=0.34$.

With our current data set, correlated fits generally give very low $p$ values. Even after drastic thinning of the data to reduce the number of small eigenvalues of the correlation matrix,  correlated fits achieve, at best, $p\sim0.01$.  The central fit shown in \figref{EMfit} is therefore an
uncorrelated fit, although we do include the correlated fits
as alternatives in the estimate of the systematic error.  We note that we have some preliminary data on additional ensembles with 
$a\!\approx\! 0.06$ fm and $a\!\approx\! 0.045$ fm, not shown in \tabref{ensembles}.   With that data, we may drop the coarsest ($a\!\approx\! 
0.12$ fm)
ensembles from the fit.  Good correlated fits then become possible, even without extreme thinning of the data.

The black lines in \figref{EMfit} show the fit after setting valence and sea masses equal,
adjusting $m_s$ to its physical value, extrapolating to the continuum, and adjusting the sea charges to
their physical values using NLO \chpt. The last adjustment vanishes identically for pions and is very small for kaons. 
From the black lines for the $\pi^+$ and $K^+$,
we subtract the corresponding results for the neutral mesons, $``\pi^0"$ and $K^0$, giving
the purple lines, whose values at the physical point for each meson (shown with
the vertical dashed-dotted lines) give the physical results.    

{\it Errors, Results, and Outlook. --} From the central fit, and using 
$(M^2_{\pi^\pm}-M^2_{\pi^0})^\gamma$ in the right-hand side of \eq{eps-def},
we find $\epsilon=0.84(5)$.  Without correction for FV effects, we found $\epsilon=0.65(7)$ \cite{EM12}.  We take 0.06, which
is 30\% of the size of the correction, as the residual FV error. One way to estimate other lattice systematic errors is to consider
a wide variety of alternative  chiral/continuum fits: different amounts of thinning of the data, different cutoffs on the highest values of 
considered, removal of various NNLO chiral and discretization terms, imposition of power-counting priors on the NNLO or NLO terms, 
and whether or not to include correlations.  
While most of these alternatives
give results fairly close to the central value, there are some outliers that also have fairly large statistical errors, especially among the 
uncorrelated fits.  Weighting the results by the inverse square of their statistical errors, the standard deviation of $\epsilon$ over all these 
alternatives is $0.07$.  The unweighted standard deviation is $0.13$, and largest amount by which differences from the central value 
exceed the statistical error is $0.21$. Since these possible estimates of the error differ greatly, we choose instead to make the estimate by 
simply replacing  $(M^2_{\pi^\pm}-M^2_{\pi^0})^{\textrm{expt}}$ instead of $(M^2_{\pi^\pm}-M^2_{\pi^0})^\gamma$; if lattice
errors were absent these two results would agree to good accuracy.  This gives a difference of $0.18$, from the central value, close to the
largest of the other estimates.
We thus find the following preliminary result:
\vspace{-1.5mm}
\begin{equation}
\epsilon = 0.84(5)_{\rm stat}(18)_{a^2}(6)_{\rm FV}\;.
\vspace{-1.5mm}
\end{equation}
Using this value for $\epsilon$, our estimate for the EM uncertainty in $m_u/m_d$ is reduced by more than
a factor of 2 from our error in \rcite{qrat}.  Using the MILC HISQ (2+1+1)-flavor QCD ensembles, as reported in \rcite{Bazavov:2014wgs}, we obtain,
preliminarily, 
\vspace{-1.5mm}
\begin{equation}
m_u/m_d = 0.4482 (48)_{\rm stat} ({}^{+\phantom{0}21}_{-115})_{a^2} (1)_{\rm FV_{QCD}} (165)_{\rm EM},
\vspace{-1.5mm}
\end{equation}
where here ``EM'' denotes all errors from EM, while ``FV$_{\rm QCD}$'' refers to finite-volume effects in the pure QCD calculation.
We are currently finishing the analysis on the additional ensembles at $a\approx 0.06$ fm and $a\approx0.045$ fm. Preliminary indications are that the lattice errors in $\epsilon$ will be significantly reduced by the inclusion of these ensembles.


{\bf Acknowledgments:} 
We thank Laurent Lellouch, Antonin Portelli, and Francesco Sanfilippo for useful discussions. 
The spectrum running was done on computers at the National Center for
Supercomputing Applications, Indiana University, the Texas Advanced
Computing Center (TACC), and the National Institute for Computational
Science (NICS). Configurations were generated with resources provided by the 
USQCD
Collaboration, the Argonne Leadership Computing Facility, and the National Energy Research Scientific 
Computing Center, which are funded by the Office of Science of the U.S.
Department of Energy; and with resources provided by the National Center for Atmospheric
Research,  NICS, the Pittsburgh Supercomputer
Center, the San Diego Supercomputer Center, and TACC,
which are funded through the National Science Foundation's XSEDE Program. This work
was supported in part by the U.S. Department of Energy
and the National Science Foundation. 
Fermilab is operated by Fermi
Research Alliance, LLC, under Contract No. DE-AC02-07CH11359 with the U.S. Department of Energy. 

\end{document}